\documentclass[aps,prl,twocolumn,groupedaddress,superscriptaddress,showpacs]{revtex4-1}
 \usepackage{graphicx}
 \usepackage{dcolumn}
 \usepackage{bm}
 \usepackage{amsmath}
 \usepackage{color}
\begin{document}

\title{Robust formation of skyrmions and topological Hall effect in epitaxial thin films of MnSi}

\author{Yufan Li}
\affiliation{ Department of Applied Physics and Quantum Phase Electronics Center (QPEC), University of Tokyo, Tokyo 113-8656, Japan }
\affiliation{ State Key Laboratory of Surface Physics and Department of Physics, Fudan University, Shanghai
200433, China }
\author{N. Kanazawa}
\affiliation{ Department of Applied Physics and Quantum Phase Electronics Center (QPEC), University of Tokyo, Tokyo 113-8656, Japan }
\author{X. Z. Yu}
\affiliation{ Cross-Correlated Materials Research Group (CMRG) and Correlated Electron Research Group (CERG), RIKEN Advanced Science Institute, Wako 351-0198, Japan}
\author{ A. Tsukazaki}
\affiliation{ Department of Applied Physics and Quantum Phase Electronics Center (QPEC), University of Tokyo, Tokyo 113-8656, Japan }
\author{  M. Kawasaki}
\affiliation{ Department of Applied Physics and Quantum Phase Electronics Center (QPEC), University of Tokyo, Tokyo 113-8656, Japan }
\affiliation{ Cross-Correlated Materials Research Group (CMRG) and Correlated Electron Research Group (CERG), RIKEN Advanced Science Institute, Wako 351-0198, Japan}
\author{ M. Ichikawa}
\affiliation{ Department of Applied Physics and Quantum Phase Electronics Center (QPEC), University of Tokyo, Tokyo 113-8656, Japan }
\author{ X. F. Jin}
\affiliation{ State Key Laboratory of Surface Physics and Department of Physics, Fudan University, Shanghai
200433, China }
\author{F. Kagawa}
\affiliation{ Department of Applied Physics and Quantum Phase Electronics Center (QPEC), University of Tokyo, Tokyo 113-8656, Japan }
\author{Y. Tokura}
\affiliation{ Department of Applied Physics and Quantum Phase Electronics Center (QPEC), University of Tokyo, Tokyo 113-8656, Japan }
\affiliation{ Cross-Correlated Materials Research Group (CMRG) and Correlated Electron Research Group (CERG), RIKEN Advanced Science Institute, Wako 351-0198, Japan}

\date{\today}

\begin{abstract}
Magneto-transport properties have been investigated for epitaxial thin films of B20-type MnSi grown on Si(111) substrates. Both Lorentz transmission electron microscopy (TEM) images and topological Hall effect (THE) clearly point to the robust formation of skyrmions over a wide temperature-magnetic field region. New features distinct from those of bulk MnSi are observed for epitaxial MnSi films: a shorter (nearly half) period of the spin helix and skyrmions, and an opposite sign of THE. These observations suggest versatile features of skyrmion-induced THE beyond the current understanding.
\end{abstract}


\maketitle

 The topological spin texture called skyrmion, in which the constituent spins point in all directions wrapping up a sphere, has recently been observed in the helimagnets with a non-centrosymmetric structure \cite{SkXMnSibulk, LorentzTEMFeCoSi, LorentzTEMFeGe, LorentzTEMMnSi, Multiferro}. This novel magnetic structure has attracted great interest not only for its rich physics related to the quantum Berry phase (skyrmion number) \cite{Ye1999,PBruno2004,THEtheoryNagaosa}, but also for its potential application. It is found that the skyrmion motion can be driven by a current density as low as $10^5$-$10^6$ A/m$^2$ \cite{motionSANS, motionTHE,motionLorentzTEM}, in contrast with the case of magnetic domain walls of conventional ferromagnets, in which the critical current density of $10^9$-$10^{11}$ A/m$^2$ is required \cite{DWscience, DWprl}. This high current-sensitivity may promise the potential application of the skyrmion in the next-generation magnetic recording technique as well as other related spintronic devices.

The skyrmion-hosting B20-type crystal structure is cubic but noncentrosymmetric and hence hosts the Dzyaloshinskii-Moriya (D-M) interaction. This antisymmetric spin exchange interaction competes with symmetric Heisenberg exchange and Zeeman interactions, producing rich magnetic structures, such as helical (proper screw) spin structure \cite{firstHelicalMnSi,AnisotropyMnSi}, conical spin structure \cite{firstHelicalMnSi, helicalSANS}, and field-induced ferromagnetism. The skyrmion spin texture, which is a superposition of three helices perpendicular to the external magnetic field $H$, is also a consequence of the competition but can be found only in quite a limited temperature-magnetic field ($T$-$H$) region (so-called $A$ phase) for bulk crystals \cite{SkXMnSibulk}. The coverage of skyrmion phase, however, turns out to be greatly enlarged in thin plates \cite{filmMC}; when $H$ is applied normal to film planes, the extended skyrmion phases have been observed by Lorentz TEM in free-standing thin plate specimens, the thickness of which are reduced to $<$~100~nm by ion-milling the bulk samples such as Fe$_{0.5}$Co$_{0.5}$Si \cite{LorentzTEMFeCoSi}, FeGe \cite{LorentzTEMFeGe}, MnSi \cite{LorentzTEMMnSi}, and insulating multiferroic Cu$_2$OSeO$_3$ \cite{Multiferro}.

In the light of both fundamental research and application, epitaxial growth of these B20 compound thin films is more beneficial, because the thickness can be better controlled, the transport properties can be precisely measured, and the desired patterning for the device fabrication is easy to implement. Despite the intensive effort in fabricating epitaxial B20 thin films, such as of MnSi \cite{Italycodepo2006,seedlayer,sandwich,ItalyTc2010,Karhu2010,Karhu2011}, no compelling evidence of the skyrmion formation has yet been obtained. A recent study on the fabrication of FeGe/Si(111) thin films has argued that the topological Hall effect (THE) exists over a wide $T$-$H$ region \cite{ChienFeGe}. The THE itself is not limited to the skyrmion spin structure alone \cite{PBruno2004,THEtheoryNagaosa} but emergent in other noncoplanar spin systems as well, therefore the one-to-one correspondence between skyrmion formation and THE in epitaxial thin films is to be evidenced. In this paper, we report the realization of high-quality epitaxial MnSi/Si(111) thin film and demonstrate skyrmion-derived THE through the combination of transport measurements and the real-space observations by Lorentz TEM.

\begin{figure}
\includegraphics[width=8.5cm]{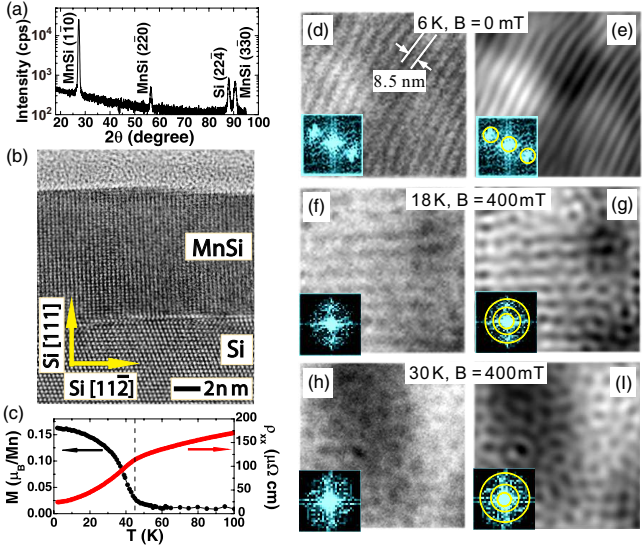}
\caption{(color online) Basic physical properties of 10~nm MnSi/Si(111) film. (a) Grazing-incident X-ray diffraction. (b) Cross-section TEM image. (c) Field-cooled magnetization measured with an in-plane external field of 50~mT $\parallel$ MnSi[11$\bar{2}$] (left axis) and $\rho_{xx}$-$T$ curve with no external field (right axis). The dashed line indicates $T_c$. (d), (f), and (h) Lorentz TEM images (over-focused) at various temperatures and magnetic fields. The corresponding fast-Fourier-transforms (FFT) are shown in the insets. (e), (g), and (l) The filtered images of the raw images (d), (f), and (h), respectively, which are obtained by shadowing the background noise and selecting magnetic reflection (circled by yellow lines in the inserted FFTs).}
\end{figure}

Our optimized scheme of the epitaxial growth of MnSi(111) thin film is to start with depositing 4 monolayer Mn at room temperatures onto Si(111)-7$\times$7 surface followed by annealing at 200~$^{\circ}$C to form a seed layer of MnSi, then depositing Mn and Si at room temperature either by co-evaporation or by repeating Mn/Si sandwich layers, and finally annealing it at 300-400~$^{\circ}$C. The single-phase nature was checked by the grazing-incident X-ray diffraction, which shows no detectable impurity phase [Fig. 1(a)]. High crystalline quality is also confirmed by cross-section TEM image [Fig. 1(b)].

The transition temperature $T_c$ of the 10 nm MnSi film is determined to be $\sim$45~K from the temperature profile of field-cooling magnetization $M$ and zero-field cooling longitudinal resistivity $\rho_{xx}$-$T$ [Fig. 1(c)]. The $T_c$ of 50~nm film is found almost identical with that of the 10~nm film. The present $T_c$ is appreciably higher than those of bulk MnSi (29.5~K) and free-standing thin film sliced off from the bulk (22.5 K for the 50 nm film), perhaps due to tensile strain induced by 3\% lattice mismatch between the epitaxial MnSi film and Si(111) substrate \cite{Karhu2010}.  Below $T_c$, a sufficiently large field aligns all the spins and turns the system into the induced ferromagnetic (spin-collinear) state. This transition manifests itself as the kink-like behavior in $M$-$H$ and $\rho_{yx}$-$H$ curves [Figs. 2(c)-(f)]. The saturated magnetization ($M_s$) at 2 K, which is derived as extrapolating $M$ from high field to zero field, is found to be $0.42 \pm 0.02$~$\mu_B$ per Mn atom, in good agreement with the reported value for bulk MnSi \cite{MnSiAHE}.

\begin{figure}
\centering
\includegraphics[width=8.5cm]{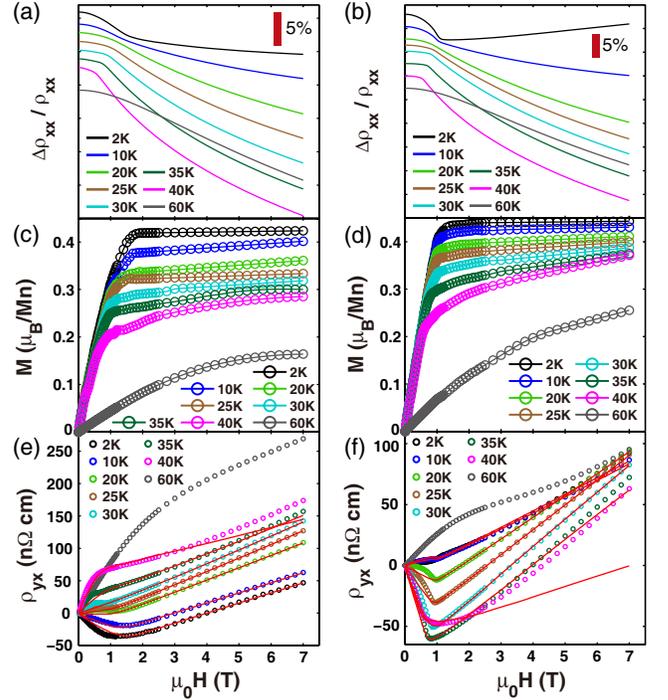}
\caption{ (color online) Magnetoresistance $\Delta$$\rho_{xx}/\rho_{xx}$, field dependence of magnetization $M$ and Hall resistivity $\rho_{yx}$ of the 10~nm [(a), (c), and (e), respectively] and 50~nm [(b), (d), and (f), respectively] films. The magnetic field was applied normal to the film plane. In (a) and (b), arbitrary offsets are added for clarity. In (c) and (d), the diamagnetic background of Si substrate estimated by the $M$-$H$ curve at 200 K is subtracted from the raw data. The red curves in (e) and (f) are the fittings for $H>H_c$ (see the text), reproducing the normal component plus anomalous Hall, $R_0H + \rho_{yx}^A$.}
\end{figure}

Figures 1(d)-(l) show representative Lorentz TEM images on the 10 nm MnSi/Si(111) film. At 6~K under zero external field, the helical structure is clearly observed as the alternating bright and dark stripes in a Lorentz TEM image [Fig. 1(d)]. The vortex-like skyrmions [bright or dark dots shown in Figs. 1(f) and (h)] can be observed at higher temperatures (18 K and 30 K) under 400 mT normal to the film. For clarity, we also present the filtered images [Figs. 1(e), 1(g), and 1(l)], which are reconstructed from the filtered magnetic reflection via the fast-Fourier-transform. Here we note that the Lorentz TEM observation reveals remarkable differences from previous results in bulk MnSi \cite{firstHelicalMnSi,helicalSANS,SkXMnSibulk} and the free-standing thin film \cite{LorentzTEMMnSi}. First, the period of helical structure ($\lambda_h$) is 8.5~nm in the 10~nm MnSi film [illustrated by the arrows in Fig. 1(d)], about half of that in the bulk specimen (18~nm). This discrepancy likely originates from different exchange coupling strengths, because $\lambda_h$ is determined by the ratio $J/D$, where $J$ and $D$ are the Heisenberg and D-M exchange couplings, respectively \cite{AnisotropyMnSi,TheoryHelicalMnSi}; the tensile strain from the substrate may alter the magnetic parameter, or substantially increase $D$. Second, the skyrmions observed in the epitaxial 10 nm film show little sign of long-range order (i.e., the hexagonal skyrmion crystal) and form a glassy state. This feature invokes the presence of disorder, which may come from slight off-stoichiometry of Mn/Si and/or lattice defects. These observations point to a sort of hierarchical nature of skyrmion formation, i.e., the local skyrmion formations themselves are robust against disorder, while their long-range crystalline packing may be prevented by the presence of disorder.

To see how the skyrmion formation affects transport properties, we looked into the field dependence of Hall resistivity $\rho_{yx}$ of the 10~nm and 50~nm films. The results are shown in Figs.~2(e) and (f) [see Figs.~4(a) and (b) for magnified views in the low-field region]. For the both films, hamp-like anomalies can be clearly seen between 15~K and 35~K when $0<H<H_c$, i.e., in accord with the $T$-$H$ region of skyrmions observed by the Lorentz TEM. We therefore ascribe this anomaly to the THE induced by the skyrmion spin texture. In general, the total Hall resistivity can be expressed as the sum of various contributions:
$\rho_{yx}=R_0H + \rho_{yx}^A + \rho_{yx}^T$,
where $R_0$ is the normal Hall coefficient, $\rho_{yx}^A$ the anomalous Hall resistivity, and $\rho_{yx}^T$ the topological Hall resistivity. In Figs.~2(e), 2(f), 4(a), and 4(b), the THE signal clearly coexists with a large background of normal Hall effect and anomalous Hall effect (AHE). To extract the THE quantitatively, $\rho_{yx}^A$ should be singled out from $\rho_{yx}$ first. Here we analyze $\rho_{yx}^A$ by employing the following scaling \cite{FeAHE,NiAHE,unifiedTheory}:
\begin{align}
\rho_{yx}^A = \alpha M\rho_{xx0}+\beta M\rho_{xx0}^2+bM\rho_{xx}^2,
\end{align}
where the terms with $\alpha$, $\beta$, and $b$ correspond to the skew scattering, the side jump, and the intrinsic contribution respectively, while $\rho_{xx0}$ stands for the residual resistivity. We derived $\rho_{yx}^A$ by extrapolating $\rho_{yx}$ in a high field region (linear fitting from 9 T to 5 T) to zero field. When $T~\rightarrow$~0~K, both $\rho_{yx}^A$ and $\rho_{xx}$ take their residual values, and thus Eq. (1) reduces into a simpler equation, $\rho_{yx0}^A/\rho_{xx0}=\alpha M_s+(\beta +b)M_s\rho_{xx0}$.
Accordingly, we show in Fig. 3(a) the result obtained at 2 K from the four 10 nm-thick samples with different $\rho_{xx0}$ values \cite{growthNote}. From the linear fitting, $\alpha M_s$ is determined to be $-3.5\times10^{-3}$. The same value seems applicable to 50~nm film as well [see Fig. 3(a)].

\begin{figure}
\centering
\includegraphics[width=6.5cm]{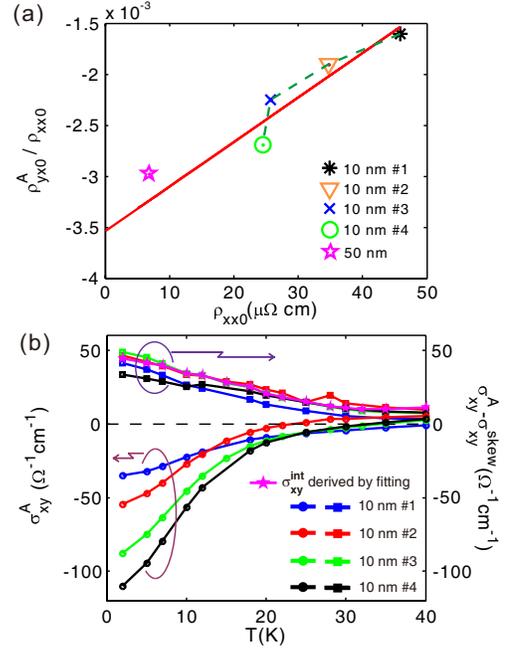}
\caption{ (color online) (a) The plot of $\rho_{yx0}^A / \rho_{xx0}$ vs $\rho_{xx0}$ obtained in the 10 nm films $\#1$-$\#4$ and the 50 nm film. The measurement was carried out at 2~K. The red line is the linear fitting to the results for the 10 nm films ($\#$1-$\#4$). (b) $\sigma_{xy}^{A}$ (circles) and $\sigma_{xy}^{A}-\sigma_{xy}^{skew}$ (squares) of the 10 nm films. $\sigma_{xy}^{int}$ (pentagrams) plotted for comparison with $\sigma_{xy}^{A}-\sigma_{xy}^{skew}$ was derived from the fitting (see the text). The superscripts are defined in the text.}
\end{figure}

In terms of conductivity, Eq. (1) can be written equivalently as \cite{FeAHE,unifiedTheory}:
\begin{align}
\sigma_{xy}^{A}=(\alpha M\sigma_{xx0}^{-1}+\beta M\sigma_{xx0}^{-2})\sigma_{xx}^2+\sigma_{xy}^{int},
\end{align}
where again the terms with $\alpha$, $\beta$ correspond to the skew scattering $\sigma_{xy}^{skew}$ and the side jump $\sigma_{xy}^{sj}$ respectively, while the last term denotes the intrinsic anomalous Hall conductivity $\sigma_{xy}^{int}$. It is straightforward to see that $\sigma_{xy}^{sj}+\sigma_{xy}^{int}=\sigma_{xy}^{A}-\alpha M\sigma_{xx0}^{-1}\sigma_{xx}^2$. We plot now in Fig.~3(b) both $\sigma_{xy}^{A}$ and $\sigma_{xy}^{A}-\sigma_{xy}^{skew}$ as a function of temperature for all the 10 nm-thick samples. It is interesting to observe that while $\sigma_{xy}^{A}$ scatters among different samples especially at low temperatures, the temperature variations of derived $\sigma_{xy}^{sj}+\sigma_{xy}^{int}$ reduce to a single universal, sample-independent curve. This feature presumably implies that the sample-dependent extrinsic term $\sigma_{xy}^{sj}$ is negligibly small, i.e., $\beta\approx 0$ \cite{NiAHE}.

The total Hall resistivity can therefore be expressed as
\begin{equation}
\rho_{yx}=R_0H + (\alpha\rho_{xx0} + b\rho_{xx}^2)M+\rho_{yx}^T.
\end{equation}
For $H>H_c$, $\rho_{yx}^T$ is supposed to be zero, because the induced ferromagnetism has no spin chirality. Hence, $R_0$ and $b$ can be determined, respectively, from the intercept and the slope of linear fitting of $(\rho_{yx}-\alpha M\rho_{xx0})/H$ vs $\rho_{xx}^2M/H$ (not shown), with the $\alpha$ value as determined above. In this analysis, we used the $\rho_{xx}$-$H$ curves shown in Figs. 2(a) and (b). The fitted results of $R_0H + (\alpha\rho_{xx0} + b\rho_{xx}^2)M$ are shown as the red solid lines in Figs.~2(e)(f) and Figs.~4(a)(b). As a crosscheck we also plotted $\sigma_{xy}^{int}$ derived from the fitting result as $\sigma_{xy}^{int}=bM$ in Fig.~3(b) and found that the temperature profile of obtained $\sigma_{xy}^{int}$ reasonably coincides with $\sigma_{xy}^{A}-\sigma_{xy}^{skew}$ as derived above. In this way, we can extract $\rho_{yx}^T$ from the difference between the total Hall resistivity $\rho_{yx}$ and the fitted curve $R_0H + (\alpha\rho_{xx0} + b\rho_{xx}^2)M$ below $H_c$.

\begin{figure}
\centering
\includegraphics[width=8.5cm]{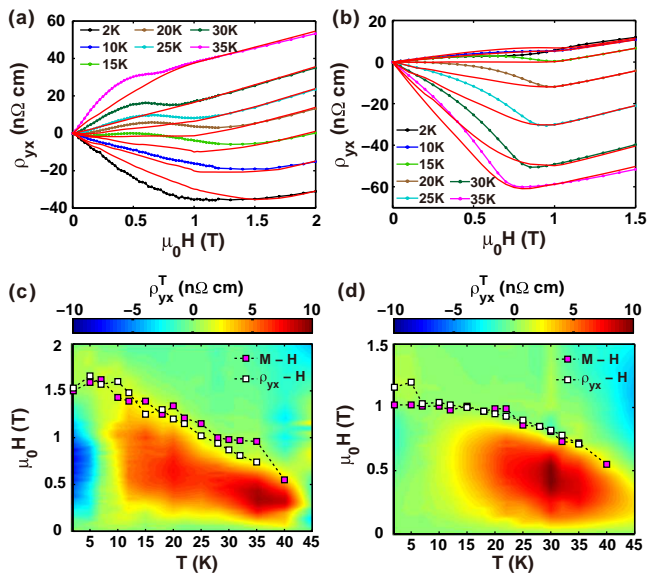}
\caption{(color online) $\rho_{yx}$ of (a) the 10 nm film and (b) the 50 nm film [magnified in the low-field region of Figs. 2(e) and (f), respectively]. The red curves are the fittings for $H>H_c$, reproducing $R_0H + \rho_{yx}^A$. (c) and (d) The contour mapping of $\rho_{yx}^T$ of (c) the 10 nm film and (d) the 50 nm film. The squares in (c) and (d) denote $H_c$ defined by the kink field in $M$-$H$ curves and $\rho_{yx}$-$H$ curves.}
\end{figure}

In the following, we discuss $\rho_{yx}^T$ in the light of the skyrmion formations. To see the global feature of $\rho_{yx}^T$, we present the contour mapping of derived $\rho_{yx}^T$ as a function of $T$ and $H$  [Figs. 4(c) and (d) for the 10 nm and 50 nm films, respectively]. A finite $\rho_{yx}^T$ is found in both films for $T>10$ K and $0<H<H_c$, consistent with the Lorentz TEM observation, which finds the skyrmion state at 18 K and 30 K under $H=400$ mT. We can thus link the onset of a positive $\rho_{yx}^T$ to the formation of the skyrmion state, and consequently Figs. 4(c) and (d) can be regarded as the magnetic phase diagrams of MnSi thin films. Obviously, the skyrmion state as evidenced by the emergence of THE is stabilized over a broad region in the $T$-$H$ plane, spanning from 10~K to $T_c$ and occupying almost all the region below $H_c$ except near zero field; this is in sharp contrast to the small $A$ phase region in bulk MnSi \cite{SkXMnSibulk,MnSiTHE}. The maximum value of $\rho_{yx}^T$ in the epitaxial thin films is about 10 n$\Omega$cm, approximately twice as large as that in bulk MnSi (-4.5 n$\Omega$cm) \cite{MnSiTHE}, albeit the signs are opposite, which will be discussed later. Furthermore, for both the 10 nm and 50 nm films, the sign of $\rho_{yx}^T$ flips from positive to negative when $T<5$ K. This behavior is more prominent in the 10 nm film [see the raw data shown in Fig. 4(a)], where the maximum magnitude of $\rho_{yx}^T$ (about -7.8 n$\Omega$cm) is comparable to the that of the positive $\rho_{yx}^T$ observed at higher temperatures. Unfortunately this temperature range ($T<5$ K) cannot be reached in our Lorentz TEM setup. The spin texture in this low temperature region requires further investigations.

The sign of $\rho_{yx}^T$ is considered to depend on the spin polarization ($P$) of charge carriers. The band structure calculation of B20 bulk MnSi indicates that the net spin polarization of the electron state near Fermi surface is very sensitive to the position of Fermi level \cite{MnSiband2004,MnSiband2008}, and the sign of $P$ may be inverted even within the accuracy of the calculation \cite{MnSiband2004}. The density of states near Fermi level is overwhelmingly contributed by $d$-electrons \cite{MnSiband2004}, which are rather localized, and the contributions of itinerant $s$-band is rather small; this feature makes the calculation of $P$ of charge carriers even more delicate. All these facts suggest that the $P$ may be affected largely by a small change in the band structure, which could be induced, for example, by the tensile strain or even by the temperature variations. The large (by half) change of the helical period in the present thin film may also signal the modification of the electronic structure. This may explain the discrepancy of the sign of $\rho_{yx}^T$ between in epitaxial MnSi film and in bulk MnSi.

In conclusion, we have succeeded in fabricating high-quality epitaxial MnSi/Si(111) thin films that host the skyrmion phase. A combination of the Lorentz TEM and measurement of topological Hall effect reveals that the skyrmion phase is extended over a much wider temperature-magnetic field range than the skyrmionic $A$ phase of bulk MnSi. The results for the 10 nm and 50 nm MnSi films show consistently that the sign of topological Hall resistivity is opposite to that in bulk MnSi, which may reflect the sign change of the conduction electron spin polarization affected by the epitaxial lattice strain, possible nonstoichiometry, and temperature-variation.

The authors wish to thank N. Nagaosa for fruitful discussions. This work was partially supported by ``KAKENHI''.(Nos. 24224009 and 24684020) from the MEXT and by the Japan Society for the Promotion of Science (JSPS) through its ``Funding Program for World-Leading Innovative R$\&$D on Science and Technology (FIRST Program)''.


\begin{thebibliography}{33}
\expandafter\ifx\csname natexlab\endcsname\relax\def\natexlab#1{#1}\fi
\expandafter\ifx\csname bibnamefont\endcsname\relax
  \def\bibnamefont#1{#1}\fi
\expandafter\ifx\csname bibfnamefont\endcsname\relax
  \def\bibfnamefont#1{#1}\fi
\expandafter\ifx\csname citenamefont\endcsname\relax
  \def\citenamefont#1{#1}\fi
\expandafter\ifx\csname url\endcsname\relax
  \def\url#1{\texttt{#1}}\fi
\expandafter\ifx\csname urlprefix\endcsname\relax\def\urlprefix{URL }\fi
\providecommand{\bibinfo}[2]{#2}
\providecommand{\eprint}[2][]{\url{#2}}

\bibitem[{\citenamefont{M\"uhlbauer et~al.}(2009)\citenamefont{M\"uhlbauer,
  Binz, Jonietz, Pfleiderer, Rosch, Neubauer, Georgii, and
  B\"oni}}]{SkXMnSibulk}
\bibinfo{author}{\bibfnamefont{S.}~\bibnamefont{M\"uhlbauer}},
  \bibinfo{author}{\bibfnamefont{B.}~\bibnamefont{Binz}},
  \bibinfo{author}{\bibfnamefont{F.}~\bibnamefont{Jonietz}},
  \bibinfo{author}{\bibfnamefont{C.}~\bibnamefont{Pfleiderer}},
  \bibinfo{author}{\bibfnamefont{A.}~\bibnamefont{Rosch}},
  \bibinfo{author}{\bibfnamefont{A.}~\bibnamefont{Neubauer}},
  \bibinfo{author}{\bibfnamefont{R.}~\bibnamefont{Georgii}}, \bibnamefont{and}
  \bibinfo{author}{\bibfnamefont{P.}~\bibnamefont{B\"oni}},
  \bibinfo{journal}{Science} \textbf{\bibinfo{volume}{323}},
  \bibinfo{pages}{915} (\bibinfo{year}{2009}).

\bibitem[{\citenamefont{Yu et~al.}(2010)\citenamefont{Yu, Onose, Kanazawa,
  Park, Han, Matsui, Nagaosa, and Tokura}}]{LorentzTEMFeCoSi}
\bibinfo{author}{\bibfnamefont{X.~Z.} \bibnamefont{Yu}},
  \bibinfo{author}{\bibfnamefont{Y.}~\bibnamefont{Onose}},
  \bibinfo{author}{\bibfnamefont{N.}~\bibnamefont{Kanazawa}},
  \bibinfo{author}{\bibfnamefont{J.~H.} \bibnamefont{Park}},
  \bibinfo{author}{\bibfnamefont{J.~H.} \bibnamefont{Han}},
  \bibinfo{author}{\bibfnamefont{Y.}~\bibnamefont{Matsui}},
  \bibinfo{author}{\bibfnamefont{N.}~\bibnamefont{Nagaosa}}, \bibnamefont{and}
  \bibinfo{author}{\bibfnamefont{Y.}~\bibnamefont{Tokura}},
  \bibinfo{journal}{Nature} \textbf{\bibinfo{volume}{465}},
  \bibinfo{pages}{901} (\bibinfo{year}{2010}).

\bibitem[{\citenamefont{Yu et~al.}(2011)\citenamefont{Yu, Kanazawa, Onose,
  Kimoto, Zhang, Ishiwata, Matsui, and Tokura}}]{LorentzTEMFeGe}
\bibinfo{author}{\bibfnamefont{X.~Z.} \bibnamefont{Yu}},
  \bibinfo{author}{\bibfnamefont{N.}~\bibnamefont{Kanazawa}},
  \bibinfo{author}{\bibfnamefont{Y.}~\bibnamefont{Onose}},
  \bibinfo{author}{\bibfnamefont{K.}~\bibnamefont{Kimoto}},
  \bibinfo{author}{\bibfnamefont{W.}~\bibnamefont{Zhang}},
  \bibinfo{author}{\bibfnamefont{S.}~\bibnamefont{Ishiwata}},
  \bibinfo{author}{\bibfnamefont{Y.}~\bibnamefont{Matsui}}, \bibnamefont{and}
  \bibinfo{author}{\bibfnamefont{Y.}~\bibnamefont{Tokura}},
  \bibinfo{journal}{Nat. Mat.} \textbf{\bibinfo{volume}{10}},
  \bibinfo{pages}{106} (\bibinfo{year}{2011}).

\bibitem[{\citenamefont{Tonomura et~al.}(2012)\citenamefont{Tonomura, Yu,
  Yanagisawa, Matsuda, Onose, Kanazawa, Park, and Tokura}}]{LorentzTEMMnSi}
\bibinfo{author}{\bibfnamefont{A.}~\bibnamefont{Tonomura}},
  \bibinfo{author}{\bibfnamefont{X.~Z.} \bibnamefont{Yu}},
  \bibinfo{author}{\bibfnamefont{K.}~\bibnamefont{Yanagisawa}},
  \bibinfo{author}{\bibfnamefont{T.}~\bibnamefont{Matsuda}},
  \bibinfo{author}{\bibfnamefont{Y.}~\bibnamefont{Onose}},
  \bibinfo{author}{\bibfnamefont{N.}~\bibnamefont{Kanazawa}},
  \bibinfo{author}{\bibfnamefont{H.~S.} \bibnamefont{Park}}, \bibnamefont{and}
  \bibinfo{author}{\bibfnamefont{Y.}~\bibnamefont{Tokura}},
  \bibinfo{journal}{Nano. Lett.} \textbf{\bibinfo{volume}{12}},
  \bibinfo{pages}{1673} (\bibinfo{year}{2012}).

\bibitem[{\citenamefont{Seki et~al.}(2012)\citenamefont{Seki, Yu, Ishiwata, and
  Tokura}}]{Multiferro}
\bibinfo{author}{\bibfnamefont{S.}~\bibnamefont{Seki}},
  \bibinfo{author}{\bibfnamefont{X.~Z.} \bibnamefont{Yu}},
  \bibinfo{author}{\bibfnamefont{S.}~\bibnamefont{Ishiwata}}, \bibnamefont{and}
  \bibinfo{author}{\bibfnamefont{Y.}~\bibnamefont{Tokura}},
  \bibinfo{journal}{Science} \textbf{\bibinfo{volume}{336}},
  \bibinfo{pages}{198} (\bibinfo{year}{2012}).

\bibitem[{\citenamefont{Ye et~al.}(1999)\citenamefont{Ye, Kim, Millis,
  Shraiman, Majumdar, and Te\ifmmode \check{s}\else
  \v{s}\fi{}anovi\ifmmode~\acute{c}\else \'{c}\fi{}}}]{Ye1999}
\bibinfo{author}{\bibfnamefont{J.}~\bibnamefont{Ye}},
  \bibinfo{author}{\bibfnamefont{Y.~B.} \bibnamefont{Kim}},
  \bibinfo{author}{\bibfnamefont{A.~J.} \bibnamefont{Millis}},
  \bibinfo{author}{\bibfnamefont{B.~I.} \bibnamefont{Shraiman}},
  \bibinfo{author}{\bibfnamefont{P.}~\bibnamefont{Majumdar}}, \bibnamefont{and}
  \bibinfo{author}{\bibfnamefont{Z.}~\bibnamefont{Te\ifmmode \check{s}\else
  \v{s}\fi{}anovi\ifmmode~\acute{c}\else \'{c}\fi{}}}, \bibinfo{journal}{Phys.
  Rev. Lett.} \textbf{\bibinfo{volume}{83}}, \bibinfo{pages}{3737}
  (\bibinfo{year}{1999}).

\bibitem[{\citenamefont{Bruno et~al.}(2004)\citenamefont{Bruno, Dugaev, and
  Taillefumier}}]{PBruno2004}
\bibinfo{author}{\bibfnamefont{P.}~\bibnamefont{Bruno}},
  \bibinfo{author}{\bibfnamefont{V.~K.} \bibnamefont{Dugaev}},
  \bibnamefont{and}
  \bibinfo{author}{\bibfnamefont{M.}~\bibnamefont{Taillefumier}},
  \bibinfo{journal}{Phys. Rev. Lett.} \textbf{\bibinfo{volume}{93}},
  \bibinfo{pages}{096806} (\bibinfo{year}{2004}).

\bibitem[{\citenamefont{Onoda et~al.}(2004)\citenamefont{Onoda, Tatara, and
  Nagaosa}}]{THEtheoryNagaosa}
\bibinfo{author}{\bibfnamefont{M.}~\bibnamefont{Onoda}},
  \bibinfo{author}{\bibfnamefont{G.}~\bibnamefont{Tatara}}, \bibnamefont{and}
  \bibinfo{author}{\bibfnamefont{N.}~\bibnamefont{Nagaosa}},
  \bibinfo{journal}{J. Phys. Soc. Jpn.} \textbf{\bibinfo{volume}{73}},
  \bibinfo{pages}{2624} (\bibinfo{year}{2004}).

\bibitem[{\citenamefont{Jonietz et~al.}(2010)\citenamefont{Jonietz,
  M\"uhlbauer, Pfleiderer, Neubauer, M\"unzer, Bauer, Adams, Georgii, B\"oni,
  Duine et~al.}}]{motionSANS}
\bibinfo{author}{\bibfnamefont{F.}~\bibnamefont{Jonietz}},
  \bibinfo{author}{\bibfnamefont{S.}~\bibnamefont{M\"uhlbauer}},
  \bibinfo{author}{\bibfnamefont{C.}~\bibnamefont{Pfleiderer}},
  \bibinfo{author}{\bibfnamefont{A.}~\bibnamefont{Neubauer}},
  \bibinfo{author}{\bibfnamefont{W.}~\bibnamefont{M\"unzer}},
  \bibinfo{author}{\bibfnamefont{A.}~\bibnamefont{Bauer}},
  \bibinfo{author}{\bibfnamefont{T.}~\bibnamefont{Adams}},
  \bibinfo{author}{\bibfnamefont{R.}~\bibnamefont{Georgii}},
  \bibinfo{author}{\bibfnamefont{P.}~\bibnamefont{B\"oni}},
  \bibinfo{author}{\bibfnamefont{R.~A.} \bibnamefont{Duine}},
  \bibnamefont{et~al.}, \bibinfo{journal}{Science}
  \textbf{\bibinfo{volume}{330}}, \bibinfo{pages}{1648} (\bibinfo{year}{2010}).

\bibitem[{\citenamefont{Schulz et~al.}(2012)\citenamefont{Schulz, Ritz, Bauer,
  Halder, Wagner, Franz, Pfleiderer, Everschor, Garst, and Rosch}}]{motionTHE}
\bibinfo{author}{\bibfnamefont{T.}~\bibnamefont{Schulz}},
  \bibinfo{author}{\bibfnamefont{R.}~\bibnamefont{Ritz}},
  \bibinfo{author}{\bibfnamefont{A.}~\bibnamefont{Bauer}},
  \bibinfo{author}{\bibfnamefont{M.}~\bibnamefont{Halder}},
  \bibinfo{author}{\bibfnamefont{M.}~\bibnamefont{Wagner}},
  \bibinfo{author}{\bibfnamefont{C.}~\bibnamefont{Franz}},
  \bibinfo{author}{\bibfnamefont{C.}~\bibnamefont{Pfleiderer}},
  \bibinfo{author}{\bibfnamefont{K.}~\bibnamefont{Everschor}},
  \bibinfo{author}{\bibfnamefont{M.}~\bibnamefont{Garst}}, \bibnamefont{and}
  \bibinfo{author}{\bibfnamefont{A.}~\bibnamefont{Rosch}},
  \bibinfo{journal}{Nat. Phys.} \textbf{\bibinfo{volume}{8}},
  \bibinfo{pages}{301} (\bibinfo{year}{2012}).

\bibitem[{\citenamefont{Yu et~al.}(2012)\citenamefont{Yu, Kanazawa, Zhang,
  Nagai, Hara, Kimoto, Matsui, Onose, and Tokura}}]{motionLorentzTEM}
\bibinfo{author}{\bibfnamefont{X.~Z.} \bibnamefont{Yu}},
  \bibinfo{author}{\bibfnamefont{N.}~\bibnamefont{Kanazawa}},
  \bibinfo{author}{\bibfnamefont{W.~Z.} \bibnamefont{Zhang}},
  \bibinfo{author}{\bibfnamefont{T.}~\bibnamefont{Nagai}},
  \bibinfo{author}{\bibfnamefont{T.}~\bibnamefont{Hara}},
  \bibinfo{author}{\bibfnamefont{K.}~\bibnamefont{Kimoto}},
  \bibinfo{author}{\bibfnamefont{Y.}~\bibnamefont{Matsui}},
  \bibinfo{author}{\bibfnamefont{Y.}~\bibnamefont{Onose}}, \bibnamefont{and}
  \bibinfo{author}{\bibfnamefont{Y.}~\bibnamefont{Tokura}},
  \bibinfo{journal}{Nat. Commun.} \textbf{\bibinfo{volume}{3}},
  \bibinfo{pages}{988} (\bibinfo{year}{2012}).

\bibitem[{\citenamefont{Myers et~al.}(1999)\citenamefont{Myers, Ralph, Katine,
  Louie, and Buhrman}}]{DWscience}
\bibinfo{author}{\bibfnamefont{E.~B.} \bibnamefont{Myers}},
  \bibinfo{author}{\bibfnamefont{D.~C.} \bibnamefont{Ralph}},
  \bibinfo{author}{\bibfnamefont{J.~A.} \bibnamefont{Katine}},
  \bibinfo{author}{\bibfnamefont{R.~N.} \bibnamefont{Louie}}, \bibnamefont{and}
  \bibinfo{author}{\bibfnamefont{R.~A.} \bibnamefont{Buhrman}},
  \bibinfo{journal}{Science} \textbf{\bibinfo{volume}{285}},
  \bibinfo{pages}{867} (\bibinfo{year}{1999}).

\bibitem[{\citenamefont{Feigenson et~al.}(2007)\citenamefont{Feigenson, Reiner,
  and Klein}}]{DWprl}
\bibinfo{author}{\bibfnamefont{M.}~\bibnamefont{Feigenson}},
  \bibinfo{author}{\bibfnamefont{J.~W.} \bibnamefont{Reiner}},
  \bibnamefont{and} \bibinfo{author}{\bibfnamefont{L.}~\bibnamefont{Klein}},
  \bibinfo{journal}{Phys. Rev. Lett.} \textbf{\bibinfo{volume}{98}},
  \bibinfo{pages}{247204} (\bibinfo{year}{2007}).

\bibitem[{\citenamefont{Ishikawa et~al.}(1976)\citenamefont{Ishikawa, Tajima,
  Bloch, and Roth}}]{firstHelicalMnSi}
\bibinfo{author}{\bibfnamefont{Y.}~\bibnamefont{Ishikawa}},
  \bibinfo{author}{\bibfnamefont{K.}~\bibnamefont{Tajima}},
  \bibinfo{author}{\bibfnamefont{D.}~\bibnamefont{Bloch}}, \bibnamefont{and}
  \bibinfo{author}{\bibfnamefont{M.}~\bibnamefont{Roth}},
  \bibinfo{journal}{Solid State Commun.} \textbf{\bibinfo{volume}{19}},
  \bibinfo{pages}{525} (\bibinfo{year}{1976}).

\bibitem[{\citenamefont{Bak and Jensen}(1980)}]{AnisotropyMnSi}
\bibinfo{author}{\bibfnamefont{P.}~\bibnamefont{Bak}} \bibnamefont{and}
  \bibinfo{author}{\bibfnamefont{M.~H.} \bibnamefont{Jensen}},
  \bibinfo{journal}{J. Phys. C} \textbf{\bibinfo{volume}{13}},
  \bibinfo{pages}{L881} (\bibinfo{year}{1980}).

\bibitem[{\citenamefont{Grigoriev et~al.}(2006)\citenamefont{Grigoriev,
  Maleyev, Okorokov, Chetverikov, B\"oni, Georgii, Lamago, Eckerlebe, and
  Pranzas}}]{helicalSANS}
\bibinfo{author}{\bibfnamefont{S.~V.} \bibnamefont{Grigoriev}},
  \bibinfo{author}{\bibfnamefont{S.~V.} \bibnamefont{Maleyev}},
  \bibinfo{author}{\bibfnamefont{A.~I.} \bibnamefont{Okorokov}},
  \bibinfo{author}{\bibfnamefont{Y.~O.} \bibnamefont{Chetverikov}},
  \bibinfo{author}{\bibfnamefont{P.}~\bibnamefont{B\"oni}},
  \bibinfo{author}{\bibfnamefont{R.}~\bibnamefont{Georgii}},
  \bibinfo{author}{\bibfnamefont{D.}~\bibnamefont{Lamago}},
  \bibinfo{author}{\bibfnamefont{H.}~\bibnamefont{Eckerlebe}},
  \bibnamefont{and} \bibinfo{author}{\bibfnamefont{K.}~\bibnamefont{Pranzas}},
  \bibinfo{journal}{Phys. Rev. B} \textbf{\bibinfo{volume}{74}},
  \bibinfo{pages}{214414} (\bibinfo{year}{2006}).

\bibitem[{\citenamefont{Yi et~al.}(2009)\citenamefont{Yi, Onoda, Nagaosa, and
  Han}}]{filmMC}
\bibinfo{author}{\bibfnamefont{S.~D.} \bibnamefont{Yi}},
  \bibinfo{author}{\bibfnamefont{S.}~\bibnamefont{Onoda}},
  \bibinfo{author}{\bibfnamefont{N.}~\bibnamefont{Nagaosa}}, \bibnamefont{and}
  \bibinfo{author}{\bibfnamefont{J.~H.} \bibnamefont{Han}},
  \bibinfo{journal}{Phys. Rev. B} \textbf{\bibinfo{volume}{80}},
  \bibinfo{pages}{054416} (\bibinfo{year}{2009}).

\bibitem[{\citenamefont{Magnano et~al.}(2006)\citenamefont{Magnano, Carleschi,
  Nicolaou, Pardini, Zangrando, and Parmigiani}}]{Italycodepo2006}
\bibinfo{author}{\bibfnamefont{E.}~\bibnamefont{Magnano}},
  \bibinfo{author}{\bibfnamefont{E.}~\bibnamefont{Carleschi}},
  \bibinfo{author}{\bibfnamefont{A.}~\bibnamefont{Nicolaou}},
  \bibinfo{author}{\bibfnamefont{T.}~\bibnamefont{Pardini}},
  \bibinfo{author}{\bibfnamefont{M.}~\bibnamefont{Zangrando}},
  \bibnamefont{and}
  \bibinfo{author}{\bibfnamefont{F.}~\bibnamefont{Parmigiani}},
  \bibinfo{journal}{Surface Science} \textbf{\bibinfo{volume}{600}},
  \bibinfo{pages}{3932} (\bibinfo{year}{2006}).

\bibitem[{\citenamefont{Higashi et~al.}(2009)\citenamefont{Higashi, Koc\'an,
  and Tochihara}}]{seedlayer}
\bibinfo{author}{\bibfnamefont{S.}~\bibnamefont{Higashi}},
  \bibinfo{author}{\bibfnamefont{P.}~\bibnamefont{Koc\'an}}, \bibnamefont{and}
  \bibinfo{author}{\bibfnamefont{H.}~\bibnamefont{Tochihara}},
  \bibinfo{journal}{Phys. Rev. B} \textbf{\bibinfo{volume}{79}},
  \bibinfo{pages}{205312} (\bibinfo{year}{2009}).

\bibitem[{\citenamefont{Azatyan et~al.}(2011)\citenamefont{Azatyan, Utas,
  Denisov, Zotov, and Saranin}}]{sandwich}
\bibinfo{author}{\bibfnamefont{S.}~\bibnamefont{Azatyan}},
  \bibinfo{author}{\bibfnamefont{O.}~\bibnamefont{Utas}},
  \bibinfo{author}{\bibfnamefont{N.}~\bibnamefont{Denisov}},
  \bibinfo{author}{\bibfnamefont{A.}~\bibnamefont{Zotov}}, \bibnamefont{and}
  \bibinfo{author}{\bibfnamefont{A.}~\bibnamefont{Saranin}},
  \bibinfo{journal}{Surface Science} \textbf{\bibinfo{volume}{605}},
  \bibinfo{pages}{289} (\bibinfo{year}{2011}).

\bibitem[{\citenamefont{Magnano et~al.}(2010)\citenamefont{Magnano, Bondino,
  Cepek, Parmigiani, and Mozzati}}]{ItalyTc2010}
\bibinfo{author}{\bibfnamefont{E.}~\bibnamefont{Magnano}},
  \bibinfo{author}{\bibfnamefont{F.}~\bibnamefont{Bondino}},
  \bibinfo{author}{\bibfnamefont{C.}~\bibnamefont{Cepek}},
  \bibinfo{author}{\bibfnamefont{F.}~\bibnamefont{Parmigiani}},
  \bibnamefont{and} \bibinfo{author}{\bibfnamefont{M.~C.}
  \bibnamefont{Mozzati}}, \bibinfo{journal}{Appl. Phys. Lett.}
  \textbf{\bibinfo{volume}{96}}, \bibinfo{pages}{152503}
  (\bibinfo{year}{2010}).

\bibitem[{\citenamefont{Karhu et~al.}(2010)\citenamefont{Karhu, Kahwaji,
  Monchesky, Parsons, Robertson, and Maunders}}]{Karhu2010}
\bibinfo{author}{\bibfnamefont{E.~A.} \bibnamefont{Karhu}},
  \bibinfo{author}{\bibfnamefont{S.}~\bibnamefont{Kahwaji}},
  \bibinfo{author}{\bibfnamefont{T.~L.} \bibnamefont{Monchesky}},
  \bibinfo{author}{\bibfnamefont{C.}~\bibnamefont{Parsons}},
  \bibinfo{author}{\bibfnamefont{M.~D.} \bibnamefont{Robertson}},
  \bibnamefont{and} \bibinfo{author}{\bibfnamefont{C.}~\bibnamefont{Maunders}},
  \bibinfo{journal}{Phys. Rev. B} \textbf{\bibinfo{volume}{82}},
  \bibinfo{pages}{184417} (\bibinfo{year}{2010}).

\bibitem[{\citenamefont{Karhu et~al.}(2011)\citenamefont{Karhu, Kahwaji,
  Robertson, Fritzsche, Kirby, Majkrzak, and Monchesky}}]{Karhu2011}
\bibinfo{author}{\bibfnamefont{E.~A.} \bibnamefont{Karhu}},
  \bibinfo{author}{\bibfnamefont{S.}~\bibnamefont{Kahwaji}},
  \bibinfo{author}{\bibfnamefont{M.~D.} \bibnamefont{Robertson}},
  \bibinfo{author}{\bibfnamefont{H.}~\bibnamefont{Fritzsche}},
  \bibinfo{author}{\bibfnamefont{B.~J.} \bibnamefont{Kirby}},
  \bibinfo{author}{\bibfnamefont{C.~F.} \bibnamefont{Majkrzak}},
  \bibnamefont{and} \bibinfo{author}{\bibfnamefont{T.~L.}
  \bibnamefont{Monchesky}}, \bibinfo{journal}{Phys. Rev. B}
  \textbf{\bibinfo{volume}{84}}, \bibinfo{pages}{060404}
  (\bibinfo{year}{2011}).

\bibitem[{\citenamefont{Huang and Chien}(2012)}]{ChienFeGe}
\bibinfo{author}{\bibfnamefont{S.~X.} \bibnamefont{Huang}} \bibnamefont{and}
  \bibinfo{author}{\bibfnamefont{C.~L.} \bibnamefont{Chien}},
  \bibinfo{journal}{Phys. Rev. Lett.} \textbf{\bibinfo{volume}{108}},
  \bibinfo{pages}{267201} (\bibinfo{year}{2012}).

\bibitem[{\citenamefont{Lee et~al.}(2007)\citenamefont{Lee, Onose, Tokura, and
  Ong}}]{MnSiAHE}
\bibinfo{author}{\bibfnamefont{M.}~\bibnamefont{Lee}},
  \bibinfo{author}{\bibfnamefont{Y.}~\bibnamefont{Onose}},
  \bibinfo{author}{\bibfnamefont{Y.}~\bibnamefont{Tokura}}, \bibnamefont{and}
  \bibinfo{author}{\bibfnamefont{N.~P.} \bibnamefont{Ong}},
  \bibinfo{journal}{Phys. Rev. B} \textbf{\bibinfo{volume}{75}},
  \bibinfo{pages}{172403} (\bibinfo{year}{2007}).

\bibitem[{\citenamefont{Nakanishi et~al.}(1980)\citenamefont{Nakanishi, Yanase,
  Hasegawa, and Kataoka}}]{TheoryHelicalMnSi}
\bibinfo{author}{\bibfnamefont{O.}~\bibnamefont{Nakanishi}},
  \bibinfo{author}{\bibfnamefont{A.}~\bibnamefont{Yanase}},
  \bibinfo{author}{\bibfnamefont{A.}~\bibnamefont{Hasegawa}}, \bibnamefont{and}
  \bibinfo{author}{\bibfnamefont{M.}~\bibnamefont{Kataoka}},
  \bibinfo{journal}{Solid State Commun.} \textbf{\bibinfo{volume}{35}},
  \bibinfo{pages}{995} (\bibinfo{year}{1980}).

\bibitem[{\citenamefont{Tian et~al.}(2009)\citenamefont{Tian, Ye, and
  Jin}}]{FeAHE}
\bibinfo{author}{\bibfnamefont{Y.}~\bibnamefont{Tian}},
  \bibinfo{author}{\bibfnamefont{L.}~\bibnamefont{Ye}}, \bibnamefont{and}
  \bibinfo{author}{\bibfnamefont{X.}~\bibnamefont{Jin}},
  \bibinfo{journal}{Phys. Rev. Lett.} \textbf{\bibinfo{volume}{103}},
  \bibinfo{pages}{087206} (\bibinfo{year}{2009}).

\bibitem[{\citenamefont{Ye et~al.}(2012)\citenamefont{Ye, Tian, Jin, and
  Xiao}}]{NiAHE}
\bibinfo{author}{\bibfnamefont{L.}~\bibnamefont{Ye}},
  \bibinfo{author}{\bibfnamefont{Y.}~\bibnamefont{Tian}},
  \bibinfo{author}{\bibfnamefont{X.}~\bibnamefont{Jin}}, \bibnamefont{and}
  \bibinfo{author}{\bibfnamefont{D.}~\bibnamefont{Xiao}},
  \bibinfo{journal}{Phys. Rev. B} \textbf{\bibinfo{volume}{85}},
  \bibinfo{pages}{220403} (\bibinfo{year}{2012}).

\bibitem[{\citenamefont{Shitade and Nagaosa}(2012)}]{unifiedTheory}
\bibinfo{author}{\bibfnamefont{A.}~\bibnamefont{Shitade}} \bibnamefont{and}
  \bibinfo{author}{\bibfnamefont{N.}~\bibnamefont{Nagaosa}},
  \bibinfo{journal}{J. Phys. Soc. Jpn.} \textbf{\bibinfo{volume}{81}},
  \bibinfo{pages}{083704} (\bibinfo{year}{2012}).

\bibitem[{gro()}]{growthNote}
\bibinfo{note}{The 10 nm-thick film $\#1$ was MBE-grown by codepositing Mn and
  Si on MnSi seed-layer at 270$^{\circ}$C. The seed-layer was prepared as
  described in the text. The 10 nm-thick films $\#2 - \#4$ were fabricated by
  preparing Mn/Si layers on the seed-layer at room temperature followed by
  annealing as described in the text, where the layers of $\#2$ and $\#3$ were
  prepared by depositing sandwich structures of repeating Mn/Si layers in
  4~ML(monolayer)/4~ML and 12~ML/12~ML respectively, and the layer of $\#4$ by
  codepositing Mn and Si simultaneously at room temperature. The 50 nm-thick
  film was fabricated in the similar way as the 10 nm-thick film $\#3$.}

\bibitem[{\citenamefont{Neubauer et~al.}(2009)\citenamefont{Neubauer,
  Pfleiderer, Binz, Rosch, Ritz, Niklowitz, and B\"oni}}]{MnSiTHE}
\bibinfo{author}{\bibfnamefont{A.}~\bibnamefont{Neubauer}},
  \bibinfo{author}{\bibfnamefont{C.}~\bibnamefont{Pfleiderer}},
  \bibinfo{author}{\bibfnamefont{B.}~\bibnamefont{Binz}},
  \bibinfo{author}{\bibfnamefont{A.}~\bibnamefont{Rosch}},
  \bibinfo{author}{\bibfnamefont{R.}~\bibnamefont{Ritz}},
  \bibinfo{author}{\bibfnamefont{P.~G.} \bibnamefont{Niklowitz}},
  \bibnamefont{and} \bibinfo{author}{\bibfnamefont{P.}~\bibnamefont{B\"oni}},
  \bibinfo{journal}{Phys. Rev. Lett.} \textbf{\bibinfo{volume}{102}},
  \bibinfo{pages}{186602} (\bibinfo{year}{2009}).

\bibitem[{\citenamefont{Jeong and Pickett}(2004)}]{MnSiband2004}
\bibinfo{author}{\bibfnamefont{T.}~\bibnamefont{Jeong}} \bibnamefont{and}
  \bibinfo{author}{\bibfnamefont{W.~E.} \bibnamefont{Pickett}},
  \bibinfo{journal}{Phys. Rev. B} \textbf{\bibinfo{volume}{70}},
  \bibinfo{pages}{075114} (\bibinfo{year}{2004}).

\bibitem[{\citenamefont{Hortamani et~al.}(2008)\citenamefont{Hortamani,
  Sandratskii, Kratzer, Mertig, and Scheffler}}]{MnSiband2008}
\bibinfo{author}{\bibfnamefont{M.}~\bibnamefont{Hortamani}},
  \bibinfo{author}{\bibfnamefont{L.}~\bibnamefont{Sandratskii}},
  \bibinfo{author}{\bibfnamefont{P.}~\bibnamefont{Kratzer}},
  \bibinfo{author}{\bibfnamefont{I.}~\bibnamefont{Mertig}}, \bibnamefont{and}
  \bibinfo{author}{\bibfnamefont{M.}~\bibnamefont{Scheffler}},
  \bibinfo{journal}{Phys. Rev. B} \textbf{\bibinfo{volume}{78}},
  \bibinfo{pages}{104402} (\bibinfo{year}{2008}).

\end{thebibliography}
\end{document}